\documentclass[twocolumn,amsmath,amssymb,aps,prb,superscriptaddress]{revtex4-2}
\usepackage{epsf}
\usepackage{graphicx}
\usepackage{sidecap}
\usepackage{soul}
\usepackage{array}
\usepackage{amsmath}
\usepackage{amssymb}
\usepackage{dcolumn}
\usepackage{epstopdf}
\usepackage{bm}
\usepackage{amsmath}
\usepackage{physics}

\usepackage{gensymb}
\usepackage{color}
\usepackage{hyperref}
\usepackage{soul}
\sethlcolor{green}
\usepackage{bbding}
\usepackage{setspace}
\hypersetup{
    colorlinks=true,
    citecolor=red,
    linkcolor=red,
    filecolor=blue,   
    urlcolor=blue,
}

\usepackage[normalem]{ulem}

\usepackage{graphicx,color}
\usepackage{amsfonts}
\usepackage[figuresright]{rotating}  
\usepackage{amssymb}
\usepackage{amsmath}

\usepackage{bbold}
\usepackage{wrapfig,lipsum,booktabs}
\usepackage{mathtools}
\usepackage{psfrag}
\usepackage{subfigure}
\usepackage{multirow}
\usepackage{tabularx}
\usepackage{textcomp}
\usepackage{bm}
\usepackage{hyperref}
\usepackage{relsize}
\hypersetup{
 pdfnewwindow=true, colorlinks=true,
 linkcolor=blue, anchorcolor=blue,
 citecolor=blue, filecolor=blue,
 menucolor=blue, urlcolor=blue}

\def\beq{\begin{eqnarray}}
\def\eeq{\end{eqnarray}}

\usepackage[dvipsnames]{xcolor}

\def \beq {\begin{equation}}
\def \eeq {\end{equation}}
\def\bibsection{\refname}
\renewcommand{\refname}{\noindent\textbf{References}}

\begin{document}

\title{Revealing the intrinsic electronic structure and complex fermiology of YRu$_2$Si$_2$ using angle-resolved photoemission spectroscopy}

\author{Anup Pradhan Sakhya} \affiliation{Department of Physics, University of Central Florida, Orlando, Florida 32816, USA}  
\author{Sabin~Regmi} \affiliation{Department of Physics, University of Central Florida, Orlando, Florida 32816, USA}
\author{Milo Sprague} \affiliation{Department of Physics, University of Central Florida, Orlando, Florida 32816, USA} 
\author{Mazharul Islam Mondal} \affiliation{Department of Physics, University of Central Florida, Orlando, Florida 32816, USA} 
\author{Iftakhar Bin Elius} \affiliation{Department of Physics, University of Central Florida, Orlando, Florida 32816, USA} 
\author{Nathan Valadez} \affiliation{Department of Physics, University of Central Florida, Orlando, Florida 32816, USA} 
\author{Andrzej~Ptok}\affiliation{Institute of Nuclear Physics, Polish Academy of Sciences, W. E. Radzikowskiego 152, PL-31342 Krak\'{o}w, Poland}
\author{Dariusz Kaczorowski}\affiliation{Institute of Low Temperature and Structure Research, Polish Academy of Sciences, ul. Okólna 2, 50-422 Wrocław, Poland}
\author{Madhab Neupane} \thanks{Corresponding author:\href{mailto:madhab.neupane@ucf.edu}{madhab.neupane@ucf.edu}}\affiliation{Department of Physics, University of Central Florida, Orlando, Florida 32816, USA}

\begin{abstract}
We performed a detailed study of the intrinsic electronic structure of YRu$_2$Si$_2$ employing angle-resolved photoemission spectroscopy (ARPES) and density-functional theory (DFT) based first-principles calculations. Electrical and magnetic measurements were conducted on well-oriented high-quality single crystals. Bulk physical measurements indicate that the compound exhibits slightly enhanced Pauli paramagnetic behavior, accompanied by electrical transport properties reminiscent of metals. Our ARPES data reveal four-fold symmetric Fermi surface with weakly-dispersing bands around the $\overline{\text{N}}$ point originating from Ru $d$ orbitals. We observed the anisotropic characteristics of the band near the $\overline{\text{N}}$ point, showing weak dispersion in the $\overline{\text{X}}$--$\overline{\text{N}}$--$\overline{\text{X}}$ direction and minimal dispersion along the $\overline{\text{N}}$--$\overline{\Gamma}$--$\overline{\text{N}}$ direction. The electronic band structure near the Fermi level is primarily governed by the Ru $d$ orbital, with minor contributions from the Y $d$ and Si $p$ orbitals. Polarization-dependent ARPES results indicate the multi-band and multi-orbital band-character of YRu$_2$Si$_2$. 
Due to the negligible correlation effect, the observed ARPES data is found to be in good agreement with the DFT results.
\end{abstract}

\maketitle

\section{Introduction}


A wide range of physical properties are displayed by compounds with AB$_2$X$_2$ stoichiometry and ThCr$_2$Si$_2$-type crystal structure. Their peculiar physical properties arise from the alternating stacks of quasi-two-dimensional A and B$_2$X$_2$ layers in this series of compounds~\cite{Hoffmann, Fujimori}. There are numerous elemental combinations for which the ThCr$_2$Si$_2$ derivatives exist. Alkali, alkaline earth, rare earth, or actinoid metals can make up the A cations, while the entire family of transition metals may occupy the B sites. The 3$^\text{rd}$, 4$^\text{th}$, and 5$^\text{th}$ main group elements can occupy the X sites~\cite{hoting}. The valence electron count, and consequently the magnetic ground states, can vary due to the flexibility in the site occupancies. Iron-based superconductors with over 450 distinct compounds~\cite{Stewart} and heavy fermions, such as YbRh$_2$Si$_2$~\cite{SteglichYb}, CeCu$_2$Si$_2$~\cite{SteglichCeCu}, and CeRu$_2$Si$_2$~\cite{BesnusCeRu} also possess the same crystal structure. 
In the context of heavy fermion systems, particularly the phases containing cerium and uranium have received a great deal of attention during the past thirty years~\cite{hoting}. CeRu$_2$Si$_2$ is a well-known non-magnetic Kondo lattice compound with a high electronic specific heat coefficient of $350$~mJ/K$^{2}$$\cdot$mol~\cite{inoue}. The $f$ electron in CeRu$_2$Si$_2$ is found to exhibit itinerant behavior~\cite{Heibl}. The electronic structure and Fermi surfaces of LaRu$_2$Si$_2$ have been thoroughly investigated through band structure calculations~\cite{yamagami}, de Haas-van Alphen effect (dHvA) measurements~\cite{Settai, Ikezawa}, and angle-resolved photoemission spectroscopy (ARPES) measurements~\cite{denlinger}. At extremely low temperatures, superconductivity is demonstrated in several AB$_2$X$_2$ phases  such as YPt$_2$Si$_2$\cite{Shelton}, YIr$_2$Si$_2$~\cite{Hirjak}, YRh$_2$Si$_2$, and YPd$_2$Si$_2$~\cite{Palstra}. Recently, the discovery of a square magnetic-skyrmion lattice in GdRu$_2$Si$_2$ which also crystallizes in the ThCr$_2$Si$_2$ structure with $I4/mmm$ space symmetry~\cite{Slaski, Heibl, SlaskiJMMM} has further generated enormous research interest as it hosts the shortest-period skyrmion lattice ever discovered and it does not possess a geometrically frustrated lattice~\cite{Khanh}. This is important because it may have applications in quantum computing and memory devices~\cite{Eremeev}. Furthermore, recent research has uncovered the presence of flat bands located at the X points of the Brillouin zone (BZ) in the unconventional superconductor YFe$_2$Ge$_2$, which shares the same ThCr$_2$Si$_2$-type crystal structure. These flat bands primarily consist of \textit{d$_{xz}$} and \textit{d$_{yz}$} orbitals~\cite{Dessau}. YRu$_2$Si$_2$ is another compound possessing the same crystal structure as CeRu$_2$Si$_2$ but lacks any 4$f$ states. While studies on the magnetoresistance and dHvA effect of YRu$_2$Si$_2$ have been reported~\cite{Settai, Ikezawa}, the material remains relatively unexplored despite the growing interest in compounds with ThCr$_2$Si$_2$-type crystal structures. 
\indent In this manuscript, we report the electronic structure of YRu$_2$Si$_2$ determined by combining ARPES and density functional theory (DFT) calculations. Bulk physical measurements revealed that the compound is a slightly enhanced Pauli paramagnet with metallic-like electrical transport properties. Our ARPES measurements indicate multiple pockets at the Fermi level (E$_F$) which is consistent with the metallic nature of this sample. We notice a flattened region of the spectra along the $\overline{\text{N}}$--$\overline{\Gamma}$--$\overline{\text{N}}$ direction, which exhibits weak dispersion along the $\overline{\text{X}}$--$\overline{\text{N}}$--$\overline{\text{X}}$ direction. This indicates the anisotropic nature of this band, which originates from Ru $d$ orbitals. We also observe a Dirac-like electron pocket at the $\overline{\text{X}}$ point.


\section{Methods}


\paragraph*{Single crystal growth and characterization.}
Single crystals of YRu$_2$Si$_2$ were grown by the Czochralski pulling technique in an ultra-pure argon atmosphere using a tetra-arc furnace from GES Corp., Japan. The elemental constituents of high chemical purity (Y: 99.97 wt. \%, Ru: 99.95 wt. \%, Si: 99.9999 wt. \%) were taken in the stoichiometric ratio. The pulling rate was 10 mm/h and the copper hearth’s rotation speed was 3 rpm. The so-prepared ingot was wrapped with Ta foil, sealed in an evacuated silica tube, and annealed at 900 \degree C for 2 weeks. The product was checked for its crystal-chemical quality by utilizing energy-dispersive X-ray (EDX) analysis, performed using an FEI scanning electron microscope and equipped with an EDAX Genesis XM4 spectrometer. Powder X-ray diffraction (XRD) was carried out on a PANanalytical X’pert Pro diffractometer with Cu K$\alpha$ radiation. The results of these XRD and EDX experiments are presented in Fig. S1 and Fig. S2, respectively in the Supplemental Material (SM) ~\cite{Supp}. Single-crystalline specimens used in the physical properties measurements were oriented by means of single-crystal Laue X-ray backscattering, implemented in a Proto LAUE-COS system. 

\paragraph*{Transport measurements.}
Electrical resistivity measurements were performed in the temperature interval $2$--$300$~K using a Quantum Design PPMS-9 platform. The electrical leads were made of $50$~$\mu$m silver wires attached to a bar-shaped specimen with silver-epoxy paste. The experiment was performed employing a standard four-point AC technique with the electric current flowing within the tetragonal plane of the YRu$_2$Si$_2$ crystal. The resistivity measured at room temperature was about $32$~$\mu$$\ohm$ cm, and that at $2$~K was equal to $0.4$~$\mu$$\ohm$ cm. The large ratio between these two values (RRR= 80) proves high quality of the sample investigated.

\paragraph*{ARPES measurements.}
ARPES measurements were performed at the Stanford Synchrotron Radiation Lightsource (SSRL), endstation 5-2. Measurements were carried out at a temperature of $15$~K. The pressure in the UHV was maintained better than $1\times10^{-10}$ Torr. The angular and energy resolutions were set better than 0.2\degree and 15 meV, respectively. Measurements were performed using photon energies in the range of $30$~eV-$90$~eV with both LH and LV polarizations.

\paragraph*{Ab initio calculations.}
DFT calculations were performed using the projector augmented-wave (PAW) potentials~\cite{bloch} implemented in the Vienna \textit{ab initio} simulation package (VASP)~\cite{kresse, kresse1,kresse2}. Calculations were made within the generalized gradient approximation (GGA) in the Perdew, Burke, and Ernzerhof (PBE) parameterization~\cite{perdew}. The energy cutoff for the plane-wave expansion was set to $400$~eV. In the calculation, we used the experimental lattice vectors as well as atomic positions~\cite{hoting}. The theoretically obtained Fermi level is overestimated with respect to the experimentally observed Fermi level by around $0.2$~eV. In our study, we present the theoretical results with the shifted Fermi level. The electronic band structure was also evaluated within {\sc Quantum ESPRESSO}~\cite{giannozzi,giannozzi1,giannozzi2} with {\sc PsLibrary}~\cite{dal}. Exact results of the electronic band structure calculation, performed for the primitive unit cell, were used to find the tight-binding model in the basis of the maximally localized Wannier orbitals~\cite{marzari,marzari1,souza}. This was performed using the {\sc Wannier90} software~\cite{mostofi,mostofi1,pizzi}. In our calculations, we used the $8 \times 8 \times 8$ full ${\bm k}$-point mesh, starting from the $p$ orbitals of Si, as well as the $d$ and $s$ orbitals for Y and Ru. Finally, the $24$-orbital tight-binding model of  YRu$_{2}$Si$_{2}$ was used to investigate the surface Green's function for the semi-infinite system~\cite{sancho}, using the {\sc WannierTools}~\cite{wu} software.

\section{Results and discussion}
\subsection{Crystal structure}
\label{sec.cry}

The crystal structure of YRu$_2$Si$_2$ is shown in Fig.~\ref{fig1}(a). The compound crystallizes in the ThCr$_{2}$Si$_{2}$-type body-centered tetragonal structure ($I4/mmm$, space group No. 139), with lattice constants $a = b = 4.1580(6)$~\AA, and $c = 9.546(2)$~\AA~\cite{hoting}. The primitive unit cell contains one Y atom at the $2a$ ($0$,$0$,$0$), two Ru atoms at $4d$ ($0$,$1/2$,$1/4$), and two Si atoms at $4e$ ($0$,$0$,$0.3684$) Wyckoff positions~\cite{hoting}. The crystal structure of this material is mainly formed by edge-sharing RuSi$_4$ tetrahedra layers in the \textit{ab} plane, which are separated by Y atoms. An exemplary Laue diffractogram is shown in the inset to Fig.~\ref{fig1}(b). Clear tetragonal symmetry pattern with well-defined  sharp spots proved high crystalline quality of the sample investigated, and comparison of the experimental data with the simulated diffractogram indicated its (001) orientation.

\begin{figure}[h]
	\includegraphics[width=8cm]{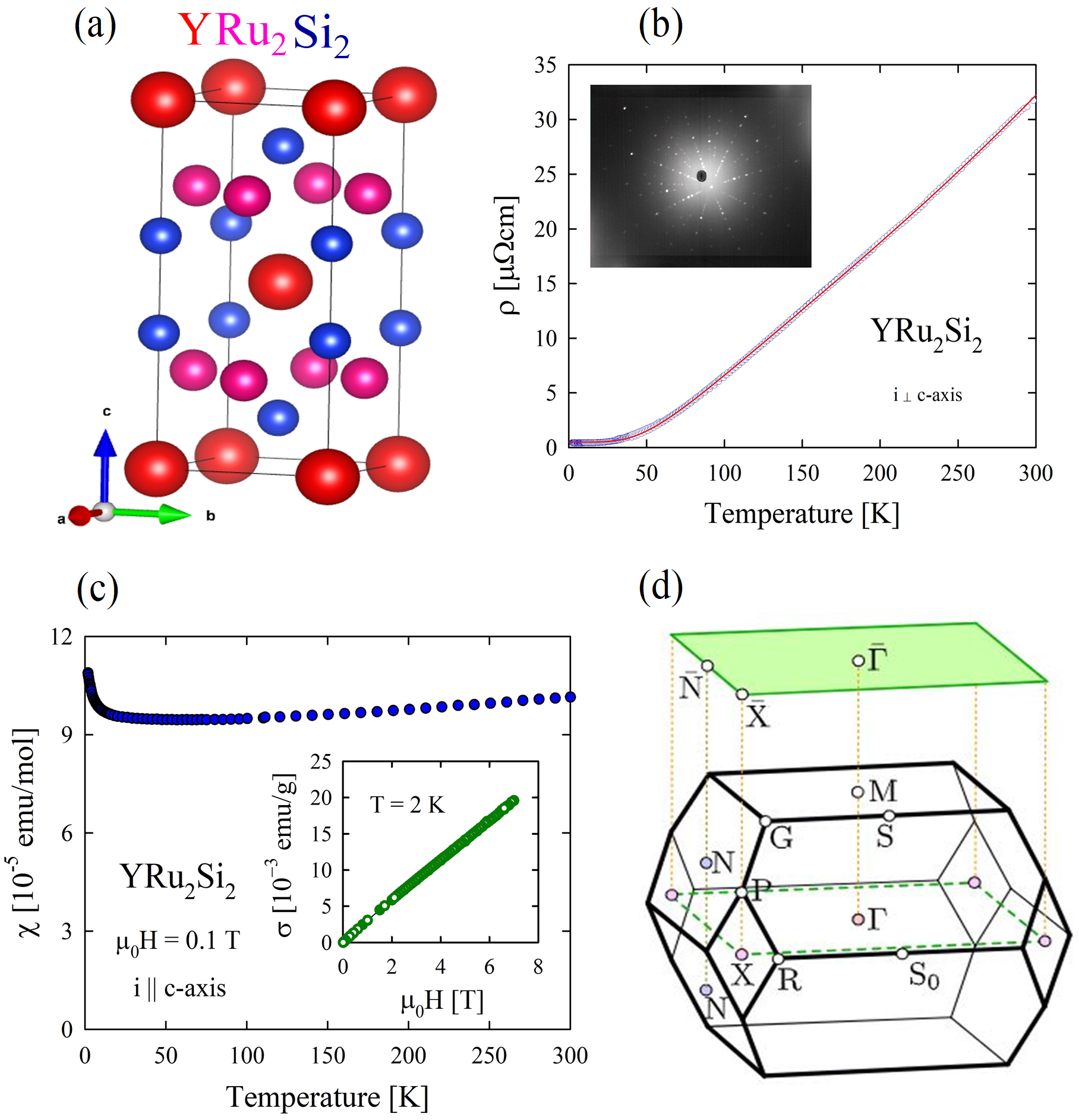} 
	\caption{Crystal structure and sample characterization of YRu$_2$Si$_2$. (a) Crystal structure of YRu$_2$Si$_2$ where the red, pink, and blue colored solid spheres denote the Y, Ru, and Si atoms, respectively. (b) Temperature dependence of the  electrical resistivity measured with electric current flowing in the tetragonal plane. Solid line represents the BGM fit discussed in the text. Inset shows the Laue diffraction pattern taken from the (001) plane. (c) Temperature dependence of the magnetic susceptibility measured in a magnetic field of 0.1 T applied along the tetragonal axis. Inset displays the magnetization isotherm measured at 2 K with increasing (solid symbols) and decreasing (open symbols) field applied along the tetragonal axis. (d) Bulk Brillouin zone (BZ) and its projection on to the (001) surface BZ. High symmetry points are marked.}
\label{fig1}
\end{figure}

\begin{figure}[h]
\includegraphics[width=\linewidth]{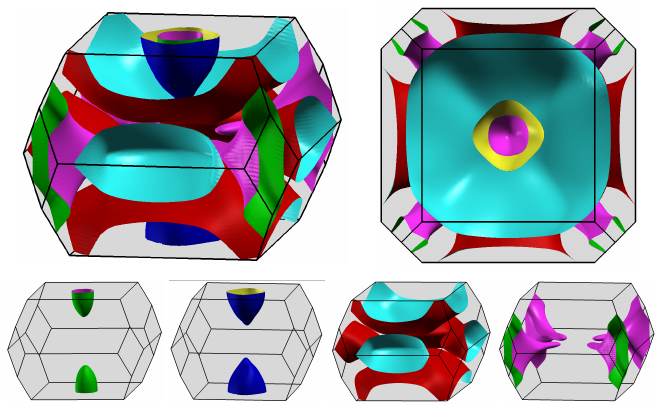}
\caption{
The Fermi surface (FS) of YRu$_{2}$Si$_{2}$ with $I4/mmm$ symmetry.
Top panels present the ``full'' FS, while bottom panels present separate pockets.
}
\label{fig.bz}
\end{figure}

\begin{figure}[h]
\includegraphics[width=\linewidth]{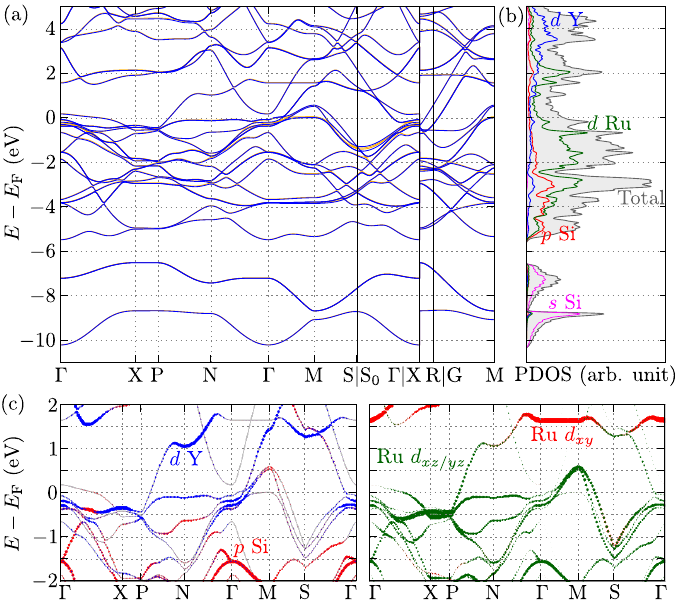}
\caption{
Theoretically obtained electronic bulk band structure (a) and corresponding density of states (b). 
The electronic band structure calculated in the absence and presence of spin-orbit coupling is represented by orange and blue lines, respectively.
The orbital projection of the band structure in the close vicinity of the Fermi level is presented in (c).
}
\label{fig.el_band}
\end{figure}

\begin{figure*} 
	\centering
	\includegraphics[width=17.5cm]{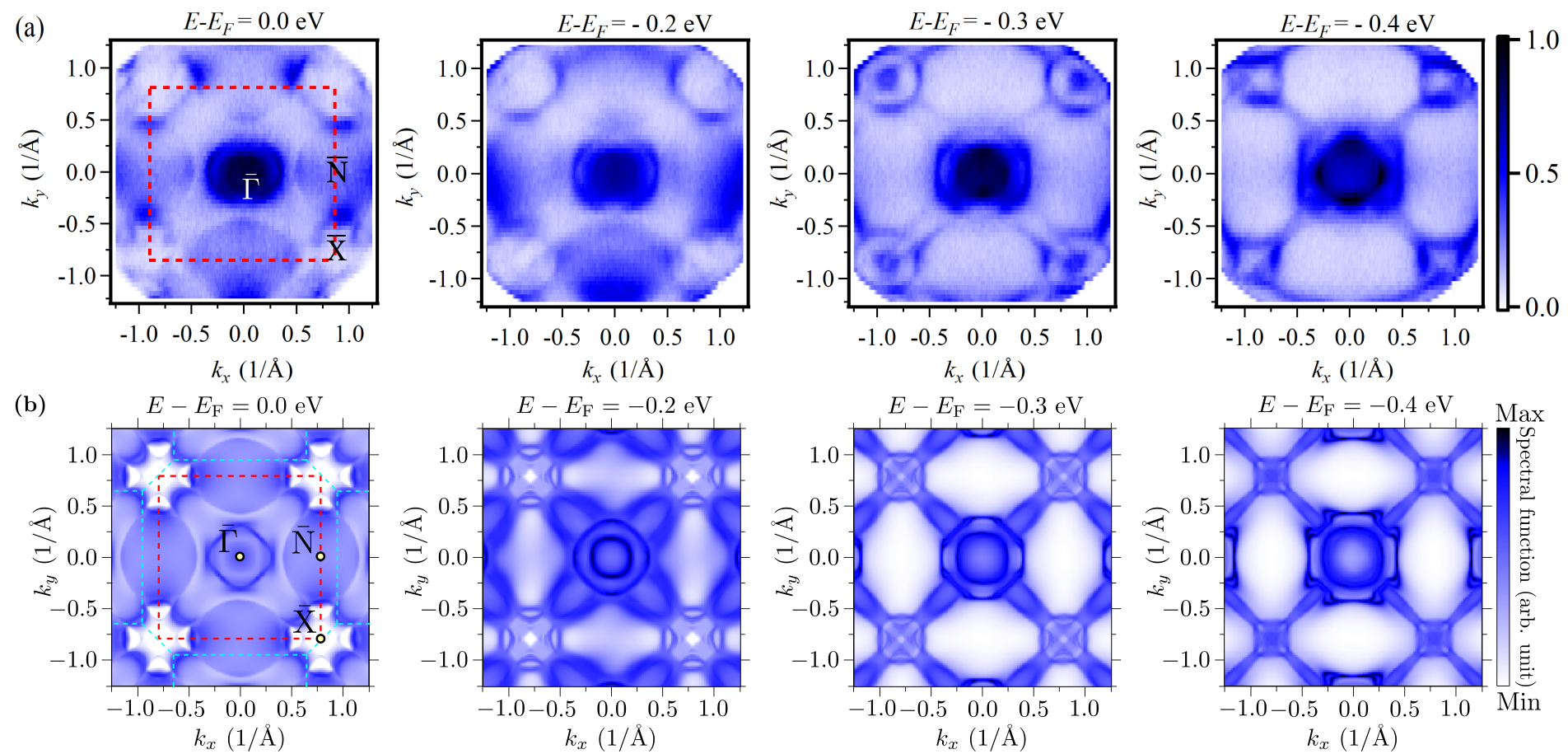}
\caption{FS and constant energy contours. (a) ARPES measured FS (first panel) and constant energy contours measured using a photon energy of $68$~eV at various binding energies as indicated on top of each plot. (b) Respective FS and constant energy contours obtained from DFT calculations.}
\label{fig2}
\end{figure*}

\subsection{Main transport properties}

The temperature variation of the resistivity ($\rho$(T)), presented in Fig.~\ref{fig1}(b), is typical for metals. The experimental  $\rho$(T) data can be well described by the Bloch-Gr{\"u}neissen-Mott (BGM) formula~\cite{Mott}:
%

\begin{eqnarray}
&\rho(T)& = \rho_{0} \\
\nonumber &+& 4RT\left(\frac{T}{\Theta_{D}}\right)^4\int_{0}^{\Theta_{D}/T}\frac{x^5 dx}{(e^x-1) (1-e^{-x})}-KT^3 ,
\end{eqnarray}
where $\rho_0$ stands for the residual resistivity due to static defects in the crystal lattice, the second term describes the phonon contribution to the total resistivity, and the third one represents \textit{s-d} interband scattering processes. The solid line in Fig.~\ref{fig1}(b) is a least-squares fit of the experimental $\rho$(T) curve to the above expression, yielding the BGM parameters: $\rho_0$= 0.4(1) $\mu$$\ohm$ cm, the Debye temperature $\Theta_D$ = 306(4) K, R = 0.09(7) $\mu\ohm$ cm/K and \textit{K} = 1.7 $\times$ 10$^{-7}$~$\mu\ohm$ cm/K$^{3}$. It is worth noting that the small value of the latter parameter, directly related to the little curvature of $\rho$(T) at high temperatures, implies that the scattering of conduction electrons via Mott-type processes in YRu$_2$Si$_2$ is almost negligible. Magnetic measurements were performed in the temperature range from 2-300 K and in magnetic fields up to 7 T using a Quantum Design MPMS-XL superconducting quantum interference device (SQUID) magnetometer. Fig.~\ref{fig1}(c) presents the temperature dependence of the molar magnetic susceptibility $\chi(T)$ =$M/H$ measured in a magnetic field of 0.1 T applied along the crystallographic \textit{c}-axis of the tetragonal unit cell of YRu$_2$Si$_2$. The compound exhibits an almost temperature-independent $\chi(T)$  of the order of 10$^{-4}$ emu/mol, implying its weakly enhanced Pauli paramagnetic character. In line with this finding, the magnetization isotherm $\sigma(H)$ taken at 2 K shows a straight-line and fully-reversible behavior up to the highest measured magnetic field of 7 T where $\sigma$ attains a small value (see the inset to Fig.~\ref{fig1}(c)). In Fig.~\ref{fig1}(d), the bulk Brillouin zone (BZ) and its projection on the (001) surface is shown where various high-symmetry points are labeled. 

\begin{figure*} 
\centering
\includegraphics[width=17.5cm]{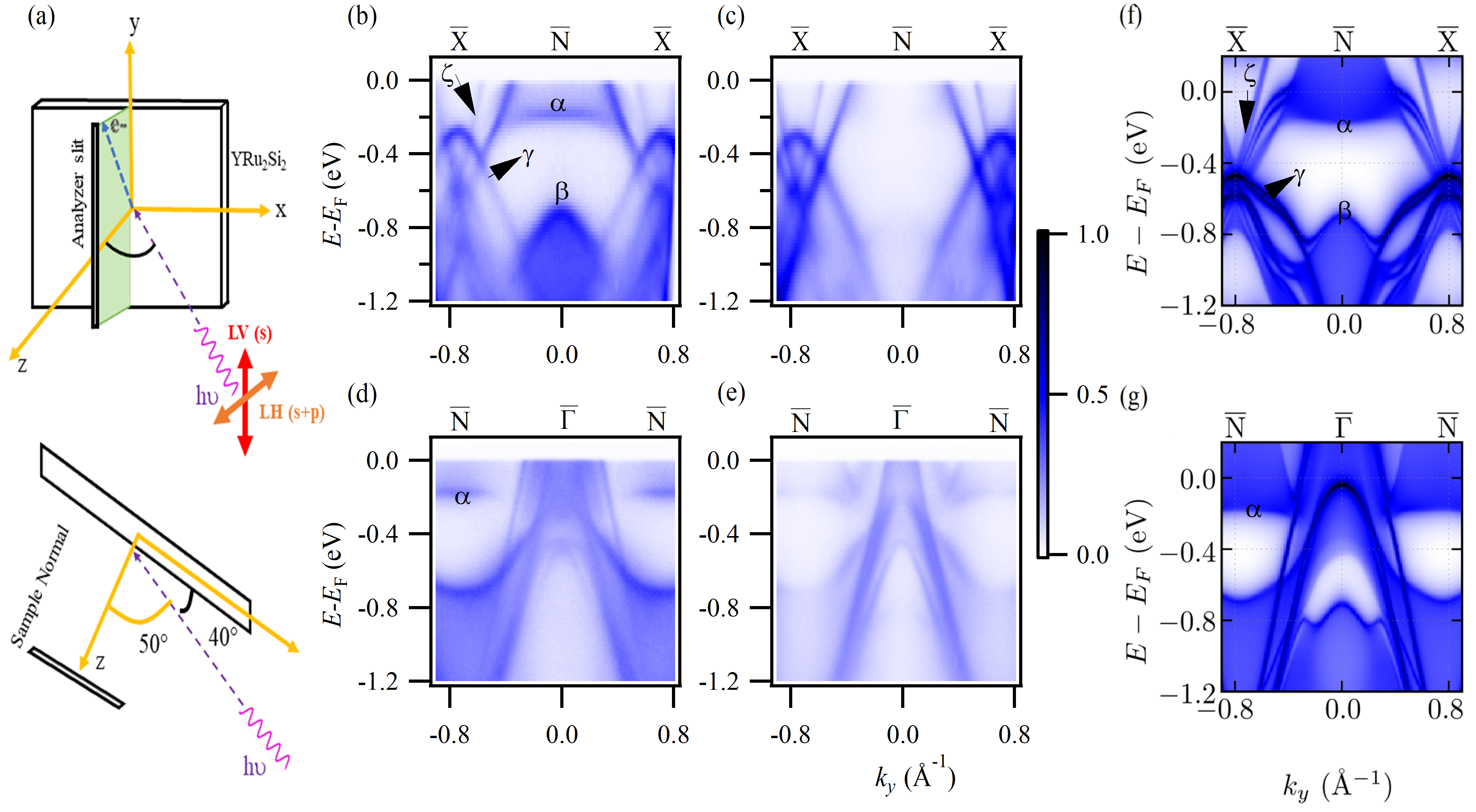} 
\vspace{-1ex}
\caption{Polarization dependent ARPES spectra of YRu$_2$Si$_2$. (a) Schematic of the experimental geometry used in the ARPES experiments. (b) Experimental band dispersion along the $\overline{\text{X}}$--$\overline{\text{N}}$--$\overline{\text{X}}$ direction using LH polarization and (c) LV polarization. (d) Experimental band dispersion along the $\overline{\text{N}}$--$\overline{\Gamma}$--$\overline{\text{N}}$ direction using LH polarization and (e) LV polarization. (f) DFT calculated band structure along the $\overline{\text{X}}$--$\overline{\text{N}}$--$\overline{\text{X}}$ direction and (g) $\overline{\text{N}}$--$\overline{\Gamma}$--$\overline{\text{N}}$ direction.
}
\label{fig3}
\end{figure*}

\begin{figure}[!b]
\centering
\includegraphics[width=0.45\textwidth]{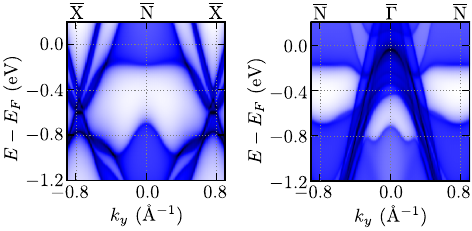} 
\vspace{-3ex}
\caption{
Theoretical results of the spectrum along $\overline{\text{X}}$--$\overline{\text{N}}$--$\overline{\text{X}}$ and $\overline{\text{N}}$--$\overline{\Gamma}$--$\overline{\text{N}}$ directions coming only from bulk states.
}
\label{fig3_bulk}
\end{figure}

\begin{figure*} 
\centering
\includegraphics[width=15cm]{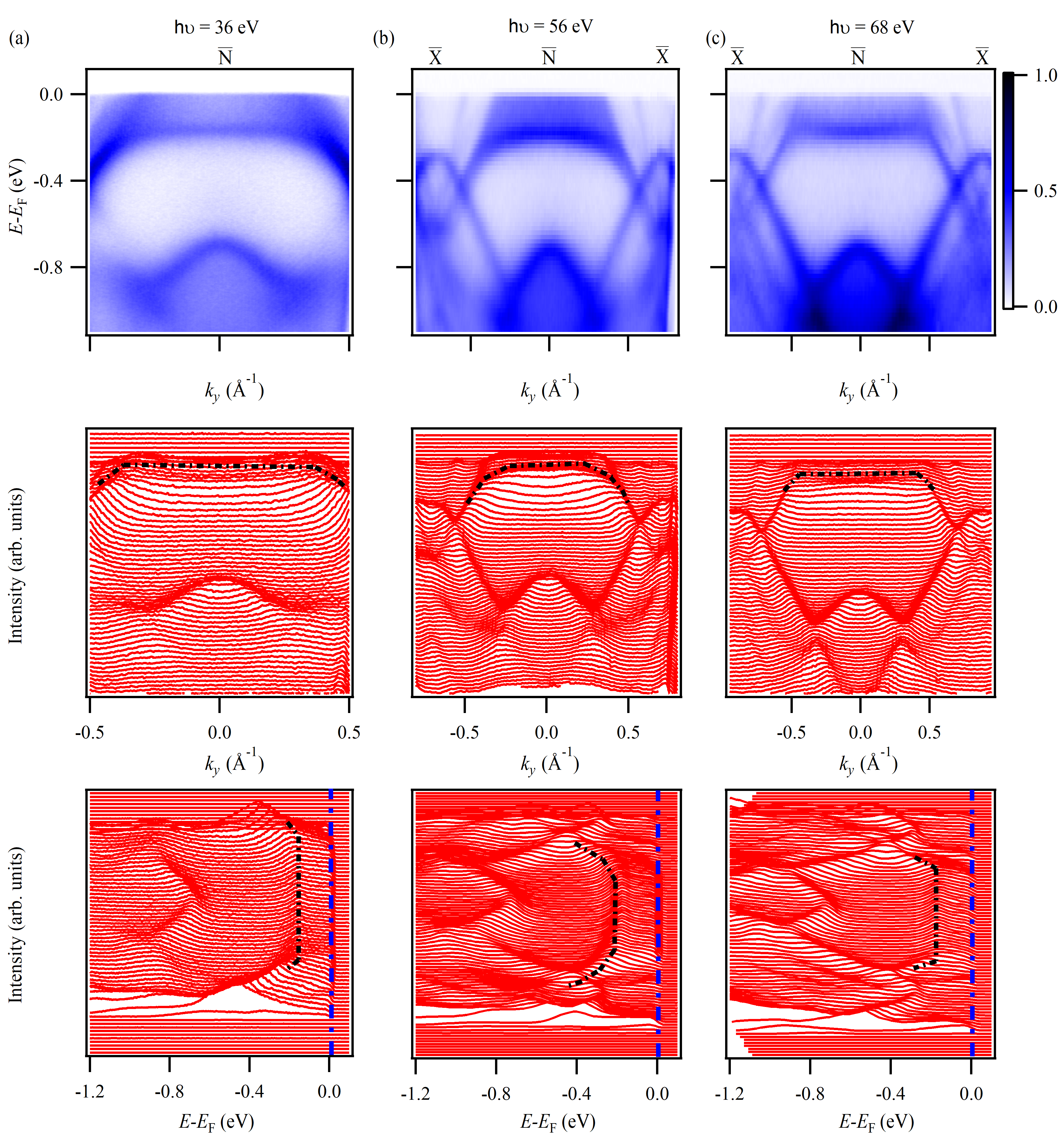} 
\vspace{-1ex}
\caption{Photon-energy dependent band dispersion along the $\overline{\text{X}}$--$\overline{\text{N}}$--$\overline{\text{X}}$ direction. Experimentally measured band dispersion along the $\overline{\text{X}}$--$\overline{\text{N}}$--$\overline{\text{X}}$ direction using photon energies of (a) 36 eV, (b) 56 eV, and (c) 68 eV, respectively. The middle and the bottom panels are the momentum distribution curves (MDCs), and the energy distribution curves (EDCs) for the respective plots in (a-c). The black dashed lines overlaid in the MDC and EDC plots are the hand-drawn curves showing the weak dispersion of the flat bands for clarity and the blue dashed lines in EDCs represent the Fermi level.}
\label{fig4}
\end{figure*}

\begin{figure*} 
\includegraphics[width=18cm]{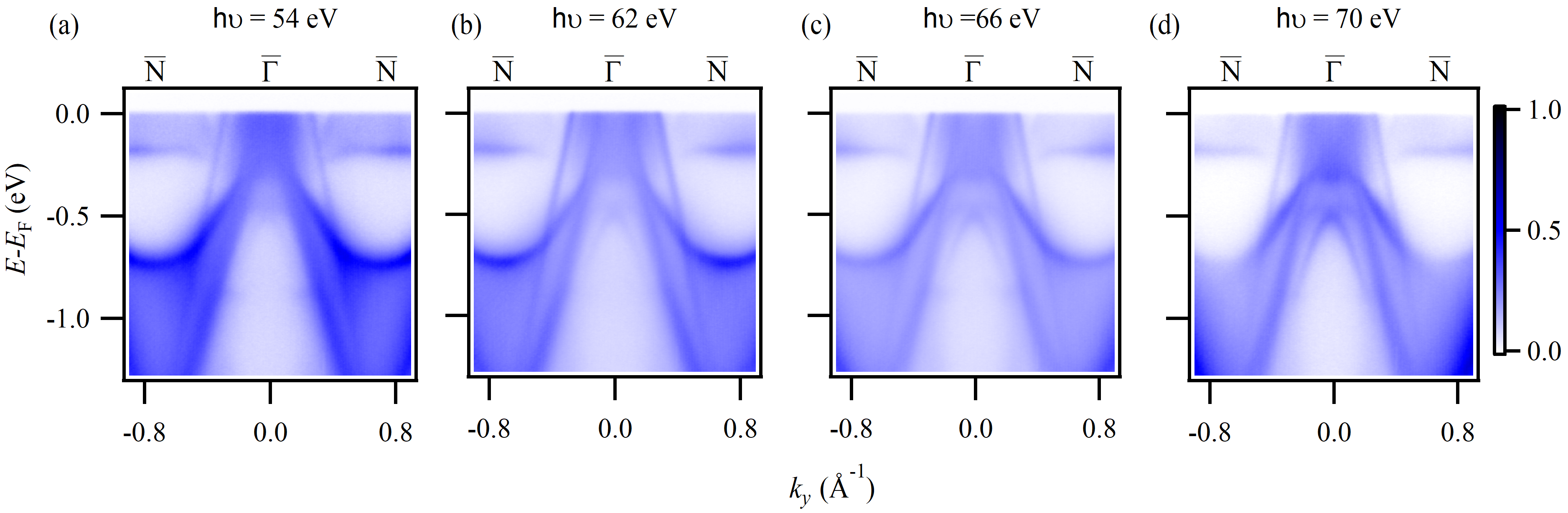}
\vspace{-1ex} 
\caption{Photon-energy dependent band dispersion along the $\overline{\text{N}}$--$\overline{\Gamma}$--$\overline{\text{N}}$ direction. (a-d) Experimentally measured band dispersion along the $\overline{\text{N}}$--$\overline{\Gamma}$--$\overline{\text{N}}$ direction at various photon energies as indicated on top of the plots using LH polarization.}
\label{fig5}
\end{figure*}

\subsection{Theoretical band structure}

The Fermi surface (FS) of YRu$_{2}$Si$_{2}$ is presented in Fig.~\ref{fig.bz}. 
The obtained FS can be directly compared with that obtained earlier for other 122 compounds.
In practice, the obtained FS has qualitatively the same shape as that reported for ThRu$_{2}$Si$_{2}$~\cite{Fujimori,Matsumoto_th}. The difference in the observed Fermi pocket sizes arises from variations in the band structure complexity and the position of the Fermi level.
In the case of YRu$_{2}$Si$_{2}$, the Fermi level is lower with respect to that reported in ThRu$_{2}$Si$_{2}$.
Such a Fermi level shift can also be a consequence of different lattice constants, stronger correlation within $f$ orbitals, or $d$-$f$ hybridization.
Nevertheless, the FS of $X$Ru$_{2}$Si$_{2}$ compounds with $f$ electrons strongly depends on the electronic configuration~\cite{denlinger}, and the similarity to ThRu$_{2}$Si$_{2}$ may be incidental. In fact, the FSs of other $X$Ru$_{2}$Si$_{2}$ compounds with $f$ electrons present much more differences~\cite{denlinger}. Similarly, the Eu-based 122 system shows Dirac electronic states \cite{Regmi, Milo}.

The shape of the FS pockets exhibits a 3D character of the electrons. 
This property is in contradiction to other 122 compounds with relatively large $c/a$ ratios. 
For example, in the case of KFe$_{2}$As$_{2}$ with $c/a \approx 3.6$, the FS takes the form of cylinders~\cite{zhao_11,watson_ba,tresca,ptok}, revealing a 2D nature of the electrons.
This is a consequence of the weak coupling between FeAs layers. 
Nonetheless, for the collapsed tetragonal phase (under pressure), $c/a$ can decrease to $2.7$. 
In this situation, the coupling between layers increases, leading to a change in the FS topology~\cite{tresca,ptok}. As a result, electrons acquire a 3D character (dispersion along $z$), while the FS is very similar to that presented in Fig.~\ref{fig.bz}. 
Moreover, such features of the FS is characteristic for 122 compounds with relatively small $c/a$, like YFe$_2$Ge$_2$~\cite{chen}. A small $c/a$ ratio, as well as the 3D character of the FS, indicates strong coupling between Ru-Si layers.

The theoretically obtained electronic band structure and density of states are presented in Fig.~\ref{fig.el_band}.
The electronic band structure possesses a relatively complex structure around the Fermi level [Fig.~\ref{fig.el_band}(a)].
The splitting of the bands introduced by spin--orbit coupling is weak, indicating a negligible role of spin--orbit coupling in the compound [cf. orange and blue lines in Fig.~\ref{fig.el_band}(a) comparing the band structures obtained without and with spin--orbit coupling].
In practice, only a small splitting of the bands along $\Gamma$--S path is visible.
The electronic density of states presented in Fig.~\ref{fig.el_band}(b) clearly shows the dominant role of the Ru $d$ orbitals. 
A small contribution from Si-$p$ and Y-$d$ orbitals around the Fermi level is also observed. The Si-$s$ states are located deep below the Fermi level, around $-9$~eV, and do not play any significant role. Such contributions are further supported by the electronic band structure projected onto the atomic orbitals [Fig.~\ref{fig.el_band}(c)].
In the close vicinity of the Fermi level, the Ru $d_{xz/yz}$ orbitals play dominant role, with a small contribution from Si $p$ and Y $d$ orbitals.

We can expect that the electronic band structure of YRu$_{2}$Si$_{2}$ will be similar to that of other $X$Ru$_{2}$Si$_{2}$ compounds.
For instance, LaRu$_{2}$Si$_{2}$ is an isostructural compound, also without $f$ electrons, and has similar lattice constants ($a = b = 4.19$~\AA\ and $c = 9.80$~\AA)~\cite{mongo_la}.
What is surprising is that the band structure of LaRu$_{2}$Si$_{2}$ possesses significant differences compared to YRu$_{2}$Si$_{2}$.
For example, we observe nearly-flat bands along the P-N-$\Gamma$ path (see Fig.~\ref{fig.el_band}), while in LaRu$_{2}$Si$_{2}$, several hole-like bands with clearly parabolic shapes are observed. The situation becomes more complex in the case of $X$Ru$_{2}$Si$_{2}$ compounds with $f$ electrons.
In such cases, the bands structures differ significantly among the family members due to the $f$ states located around the Fermi level~\cite{denlinger}.


\subsection{The ARPES measurements}

Next, we have performed ARPES measurements to reveal the electronic structure of YRu$_2$Si$_2$. Fig.~\ref{fig2}(a) shows the FS and constant energy contours (CECs) at various binding energies using a photon energy of 68 eV. Experimental FS maps and CECs are presented in Fig.~\ref{fig2}(a) and the DFT calculated FS and CECs are presented in Fig.~\ref{fig2}(b). Both experimental and theoretical FSs are provided with BZs marked with high-symmetry points. Multiple pockets are observed at the FS, which indicates the complex fermiology of this material. We observe multiple circular-shaped pockets at the $\overline{\Gamma}$ point, an ellipsoidal pocket at the $\overline{\text{N}}$ point and a pocket at the $\overline{\text{X}}$ point. The energy contours presented in the right hand side of Fig.~\ref{fig2}(a) and Fig.~\ref{fig2}(b) delineate how the band dispersions evolve with binding energies. The energy pockets at the $\overline{\Gamma}$ and the $\overline{\text{N}}$ points grow bigger in size with the increase in binding energy, indicating the hole-like nature of the bands whereas the pocket at the $\overline{\text{X}}$ point decreases in size indicating the electron-like nature. Several bulk pockets appear at higher binding energies, which can be clearly visualized from the CEC plots at binding energies of 200 meV, 300 meV, and 400 meV. The experimental FS and the CECs of the material are quite well reproduced from the DFT calculations as shown in Fig.~\ref{fig2}(b). The exception of some pockets such as the rhombus pocket at the $\overline{\Gamma}$ point can be explained due to its polarization dependence which can be seen from the polarization-dependent FS maps presented in Figs. S3-S4 in the SM~\cite{Supp}. 

While there have not been any electronic structure investigations utilizing DFT or ARPES for YRu$_2$Si$_2$, dHvA measurements have been reported~\cite{Settai, Ikezawa}. In contrast, LaRu$_2$Si$_2$, which shares a similar crystal structure as discussed previously, has undergone extensive examination via both dHvA and ARPES~\cite{Settai, Ikezawa, yamagami, denlinger}. Therefore, we compare our ARPES findings with both dHvA for LaRu$_2$Si$_2$ and YRu$_2$Si$_2$, and ARPES with LaRu$_2$Si$_2$ in the subsequent statements. 
The FS of YRu$_2$Si$_2$, as observed, shows significant congruence with the dHvA measurements~\cite{Settai, Ikezawa}, and ARPES measurements~\cite{denlinger}, especially concerning the pocket at the $\overline{\text{N}}$ point. A large, distinct closed-hole FS with an ellipsoidal shape at the Z point, corresponding to the $\overline{\text{N}}$ point in our ARPES data, was observed in both LaRu$_2$Si$_2$ and YRu$_2$Si$_2$ from both  dHvA and ARPES measurements~\cite{Settai, Ikezawa, yamagami, denlinger}. Additionally, three hole pockets were observed at the Z point for LaRu$_2$Si$_2$ from both dHvA and ARPES measurements, whereas one hole pocket was observed for YRu$_2$Si$_2$~\cite{Settai, Ikezawa, yamagami, denlinger}. Upon comparison, we confirm a large ellipsoidal hole pocket centered at the $\overline{\text{N}}$ point, which aligns well with the reported dHvA and ARPES measurements. However, discrepancies arise as additional hole pockets at the $\overline{\text{N}}$ point, resembling a rugby ball shape, were not detected in our ARPES data. Additionally, our ARPES data reveals multiple hole-like pockets at the $\overline{\Gamma}$ point, particularly emphasized in LH polarized data compared to LV polarized data (see Fig.~\ref{fig3}) which were not observed in dHvA measurements and ARPES measurements~\cite{Settai, Ikezawa, yamagami, denlinger}.

To examine the details of the band structure, we present the ARPES measured band dispersion along the $\overline{\text{X}}$--$\overline{\text{N}}$--$\overline{\text{X}}$ and the $\overline{\text{N}}$--$\overline{\Gamma}$--$\overline{\text{N}}$ direction as shown in Fig.~\ref{fig3}. We performed polarization-dependent ARPES measurements to investigate the potential multi-orbital characteristics of YRu$_2$Si$_2$. The experimental geometry of the polarization dependent ARPES is presented in Fig.~\ref{fig3}(a). ARPES provides a unique opportunity to directly examine the orbital texture of \textit{k}-states with different symmetries by using photon polarization selection rules~\cite{Damascelli}. The intensity (I) measured in ARPES experiments is strongly dependent on the transition matrix elements of the photoemission process and can be expressed as \mbox{I $\propto$ $\bra{\Psi_f}\vec{A} \cdot \vec{p}\ket{\Psi_i}$}, where $\vec{A}$ is the electromagnetic gauge and $\vec{p}$ is the electron momentum. $\vec{A}$ exhibits the same spatial mirror symmetry as the electric field $\vec{E}$ of the incident polarized photon beam while $\ket{\Psi_i}$ and $\ket{\Psi_f}$represent the electron wavefunction in the solid (initial state) and the wave function of the excited photoelectron (final state), respectively~\cite{Hufner, Himpsel}. The  incident beam and the normal to the sample surface together form a mirror plane. Given that the final state $\ket{\Psi_f}$ of photoelectrons can be approximated by a plane wave with its wavevector in the mirror plane, it is always even concerning the mirror plane~\cite{Hufner, Himpsel}.

The measurements have been performed for a photon energy of 62 eV using both LH and LV polarizations. We observe multiple bands crossing the E$_F$. The electron-like pocket at the $\overline{\text{X}}$ point as discussed above interestingly reveals a Dirac-like band (denoted as $\zeta$) which is marked by black arrow in Fig.~\ref{fig3}(b). This Dirac-like state seems to be quite intense in both LH and LV polarization measurements as can be seen in Fig.~\ref{fig3}(b) and~Fig.~\ref{fig3}(c) (see also Fig. S5 in the SM~\cite{Supp}). We also observe a flattened region of the spectra around the $\overline{\text{N}}$ point (denoted as $\alpha$) which is found to be very sensitive to polarization measurement as this band is intense when measured with LH polarization and strongly suppressed when measured with LV polarization [see Fig.~\ref{fig3}(b)-(e)]. Another hole-like band at around 800 meV below the E$_F$ (denoted as $\beta$) is found to be very sensitive to polarization measurements as the intensity of this band is strong in LH polarization, however, it is almost absent in LV polarization [see Fig.~\ref{fig3}(b)-(c)].

In the LH (or LV) geometry $\vec{A} \cdot \vec{p}$ is odd (or even) with respect to the mirror plane. Hence, taking into account the spatial symmetry of the Ru 3$d$ orbitals, when the analyzer slit aligns with the high-symmetry directions as presented in Fig.~\ref{fig3}, the photoemission intensity of specific even (or odd) components of a band can only be detected with LH (or LV) polarized light. For instance, concerning the \textit{yz}-mirror plane, LV excites even orbitals (\textit{$d_{x^{2}-y^{2}}$}, \textit{$d_{z^{2}}$}, and \textit{d$_{yz}$}) and suppresses odd orbitals (\textit{d$_{xy}$}, \textit{d$_{xz}$}), respectively. Based on these selection rules and the orbital character of the bands obtained from DFT calculations [see Fig.~\ref{fig.el_band}(c)] the flattened region of the spectra around the $\overline{\text{N}}$ high-symmetry point (denoted by $\alpha$) is strongly dominated by Ru $d_{xz}$ orbitals. Additionally, the band $\beta$ originates from Ru \textit{d$_{xz}$} orbitals. The DFT calculated surface and bulk spectrum is presented in Fig.~\ref{fig3}(f) and~\ref{fig3}(g) and is in good agreement with the observed ARPES results where the flattened region of the spectra around the $\overline{\text{N}}$ point and the Dirac-like band can be seen clearly. The above qualitative arguments obtained from polarization-dependent ARPES results, consistent with our DFT calculations allow us to delineate the contribution from different orbitals to the electronic band structure of this material. We have also presented the DFT calculated bulk spectrum along the $\overline{\text{X}}$--$\overline{\text{N}}$--$\overline{\text{X}}$ and $\overline{\text{N}}$--$\overline{\Gamma}$--$\overline{\text{N}}$ directions, including only the bulk states, as shown in Fig.~\ref{fig3_bulk}. By comparing Fig.~\ref{fig3_bulk} with Fig.~\ref{fig3}(f,g), we observe that the $\gamma$ band is not a bulk band but originates from surface states.


	
To further examine the dimensionality of these bands, we performed photon energy dependent APRES measurements along the $\overline{\text{X}}$--$\overline{\text{N}}$--$\overline{\text{X}}$ direction using photon energies of 36 eV, 56 eV, and 68 eV using LH polarization as shown in Fig.~\ref{fig4}. The flattened region of the spectra around the $\overline{\text{N}}$ point seems to weakly disperse with the change in photon energy. The momentum distribution curves (MDCs) and the energy distribution curves (EDCs) are presented in the middle and the bottom panel of Fig.~\ref{fig4}(a-c), corresponding to their respective photon energies. The black dotted lines in both the MDCs and the EDCs clearly reveal the weak dispersion of the band around the $\overline{\text{N}}$ high-symmetry point.
In Fig.~\ref{fig5} we present the photon-energy dependent bands along the $\overline{\text{N}}$--$\overline{\Gamma}$--$\overline{\text{N}}$ direction using LH polarization. The band observed at the $\overline{\text{N}}$ point seem to show no dispersion even with the change in photon energy from 54 eV to 70 eV. This is in contrast to the weak-dispersion along the $\overline{\text{X}}$--$\overline{\text{N}}$--$\overline{\text{X}}$ direction indicating the anisotropic dispersion of this Ru $d$ dominated band around the $\overline{\text{N}}$ point. The slight variation in intensity is possibly due to the photoemission matrix element effect. 

\section{Conclusions}

In summary, we synthesized high-quality single crystals of YRu$_2$Si$_2$ and characterized them using various bulk measurement techniques such as X-ray diffraction, Laue diffraction, magnetization and resistivity measurements. 
YRu$_2$Si$_2$ crystallizes in the ThCr$_{2}$Si$_{2}$ family of materials, belonging to the space group $I4/mmm$. 
Electrical resistivity measurements indicate its metallic behavior, whereas magnetization measurements exhibit slightly enhanced Pauli paramagnetic behavior.
We reproduce the experimentally obtained ARPES spectra within the DFT calculations. The excellent agreement between the experimental results and the theoretically obtained spectra indicates a weak role of the correlation effects, in contrast to the iron-based 122 compounds~\cite{borisenko.fe}.
We found the anisotropic nature of the band around the $\overline{\text{N}}$ point as it shows weak dispersion along the $\overline{\text{X}}$--$\overline{\text{N}}$--$\overline{\text{X}}$ direction and negligible dispersion along the $\overline{\text{N}}$--$\overline{\Gamma}$--$\overline{\text{N}}$ direction. 
We also observe a Dirac-like state at the $\overline{\text{X}}$ point, with a nearly-linear crossing of the non-topological surface states. 
Additionally, polarization-dependent ARPES results suggest the multi-band and multi-orbital nature of YRu$_2$Si$_2$. 
The electronic band structure around the Fermi level is dominated by Ru $d$ orbitals, with small contribution of Y $d$ and Si $p$ orbitals.

\begin{acknowledgments}
M.N. acknowledges the support from the Air Force Office of Scientific Research MURI (Grant No. FA9550-20-1-0322), and the National Science Foundation (NSF) CAREER award DMR-1847962.  A.P. acknowledges the support by National Science Centre (NCN, Poland) under Projects No. 2021/43/B/ST3/02166. The use of Stanford Synchrotron Radiation Lightsource (SSRL) in SLAC National Accelerator Laboratory is supported by the U.S. Department of Energy, Office of Science, Office of Basic Energy Sciences under Contract No. DE-AC02-76SF00515. We thank Makoto Hashimoto and Donghui Lu for the beamline assistance at SSRL endstation 5-2. Some figures in this work were rendered using {\sc Vesta}~\cite{momma} and {\sc xCrysDen}~\cite{kokalj} softwares.
\end{acknowledgments}


 \vspace{1cm}


\vspace{2ex}



\raggedbottom
\clearpage
\pagebreak{}

\clearpage
\widetext
\begin{center}
\textbf{\large Supplemental Materials for \\~\\Revealing the intrinsic electronic structure and complex fermiology of YRu$_2$Si$_2$ using angle-resolved photoemission spectroscopy}
\setcounter{equation}{0}
\setcounter{figure}{0}
\setcounter{table}{0}
\setcounter{page}{1}
\makeatletter
\renewcommand{\theequation}{S\arabic{equation}}
\renewcommand{\figurename}{{Fig.}}
\renewcommand{\thefigure}{{{S\arabic{figure}}}}
\renewcommand{\bibnumfmt}[1]{[#1]}
\renewcommand{\citenumfont}[1]{#1}
\renewcommand{\tablename}{Supplementary Table}
\renewcommand{\thetable}{\arabic{table}}
\def\bibsection{\refname}
\renewcommand{\refname}{\noindent\textbf{Supplementary References}}

\begin{center}
\textbf{Characterization of single crystals.}
\end{center}
The Czochralski-grown single crystal of YRu$_2$Si$_2$ was a rod of approximately 4 mm in diameter and 10 mm in length. The XRD pattern shown in Fig. S1 was easily indexed assuming the tetragonal ThCr$_2$Si$_2$-type crystal structure (space group $I4/mmm$). The EDX results (see Fig. S2) proved single-phase character of the sample examined with the chemical composition very close to the nominal one.

\begin{figure*} [h!]
	\includegraphics[width=10cm]{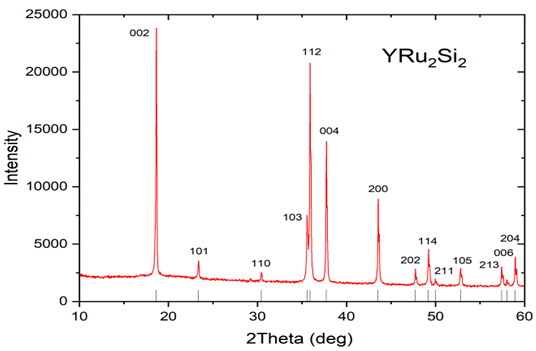} 
    \vspace{-1ex}
	\caption{Powder X-ray diffraction pattern of powdered single crystal of YRu$_2$Si$_2$. The Bragg peaks are labeled with (hkl) Miller indices.}
\label{fig1}
\end{figure*}

\begin{figure*} [h!]
\centering
\includegraphics[width=12cm]{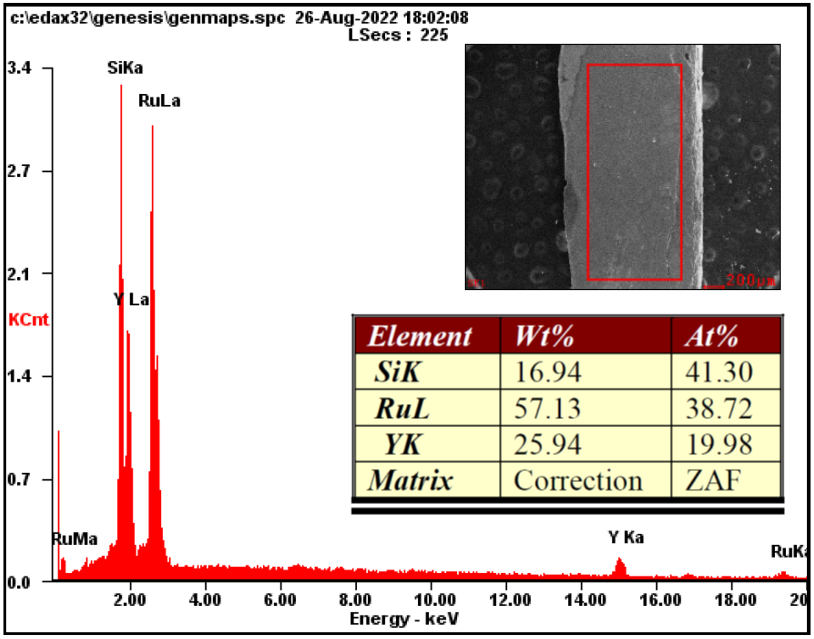}
\caption{The EDX results obtained for single-crystalline YRu$_2$Si$_2$.}
\label{figS2}
\end{figure*}

\begin{figure*}
\centering
\includegraphics[width=10cm]{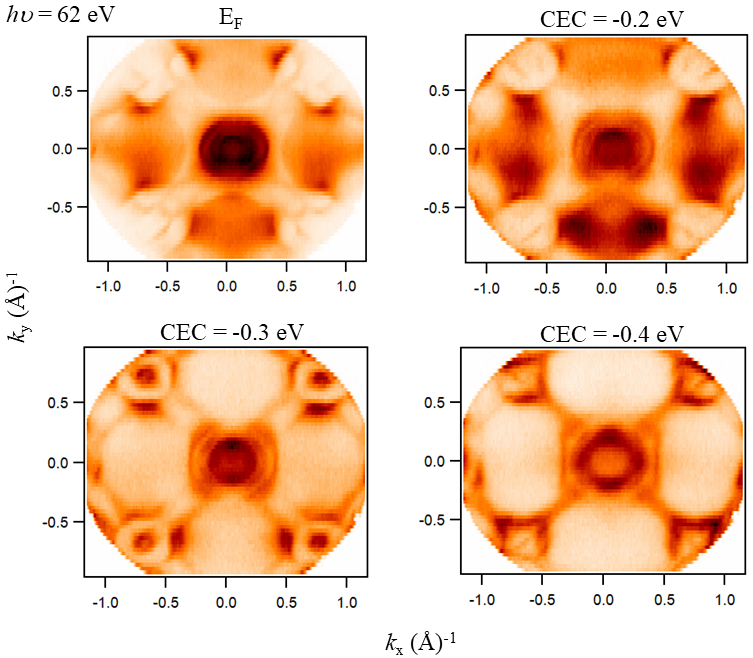} 
\caption{{ARPES measured Fermi surface and constant energy contours as noted on top of each plot.} The measurements have been performed using a photon energy of 62 eV using LH polarization. The experimental data were taken at the SSRL beamline 5-2 at a temperature of 15 K.}
\label{figS3}
\end{figure*}

\begin{figure*}
\centering
\includegraphics[width=10cm]{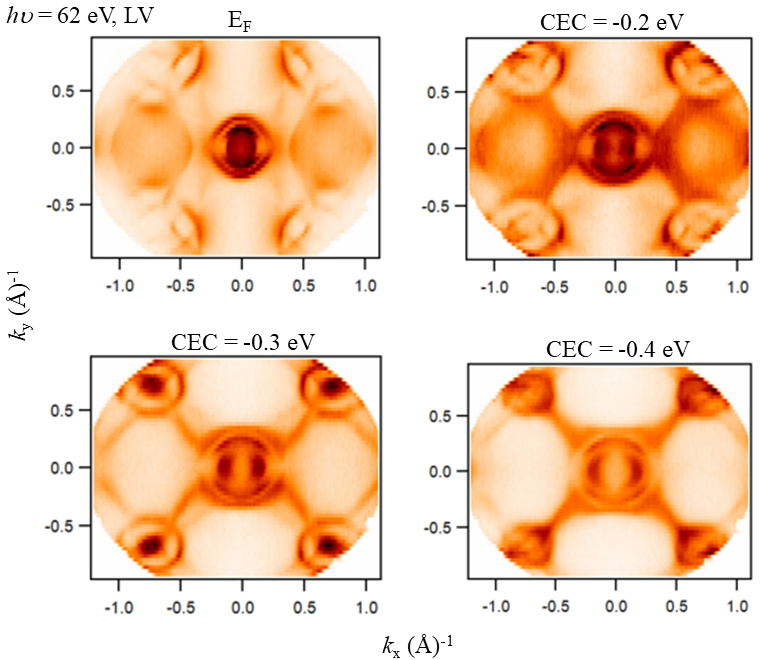} 
\caption{{ARPES measured Fermi surface and constant energy contours as noted on top of each plot.} The measurements have been performed using a photon energy of 62 eV using LV polarization. The experimental data were taken at the SSRL beamline 5-2 at a temperature of 15 K.}
\label{figS4}
\end{figure*}
\vspace{4ex}

\vspace{4ex}

\begin{center}
\textbf{Fermi surface and constant energy contour maps.}
\end{center}
\noindent  In Fig. S3 and Fig. S4, we have presented the Fermi surface and constant energy contour maps measured at a photon energy of 62 eV using LH and LV polarizations, respectively. The obtained Fermi surface shows complex metallic band structures. The hole-like nature of the pocket can be seen at the $\overline{\Gamma}$ point and the $\overline{\text{N}}$ point, whereas an electron-like pocket can be observed at the $\overline{\text{X}}$ point. The obtained Fermi surface and the constant energy contours are in excellent agreement with the theoretical calculations of the Fermi surface as discussed in Fig. 4b in the main text. Slight discrepancies in theoretical calculations with the measured ARPES data is due to polarization dependence which can be clearly seen in the Fermi surface and the constant energy contours as presented in Fig. S3 and Fig. S4.\newline

\begin{center}
\textbf{Dirac-like state at the $\overline{\text{X}}$ high-symmetry points.}
\end{center}
\noindent To better visualize the Dirac-like state at the  $\overline{\text{X}}$ high-symmetry points, we have plotted the band dispersion in a narrow range as shown in Fig. S5. The Dirac-like crossing at the $\overline{\text{X}}$ can be seen clearly. Additionally, the two-dimensional bulk Fermi surface contour highlighting various high-symmetry directions are also plotted in Fig. S6. \\

\begin{figure*}
\centering
\includegraphics[width=14cm]{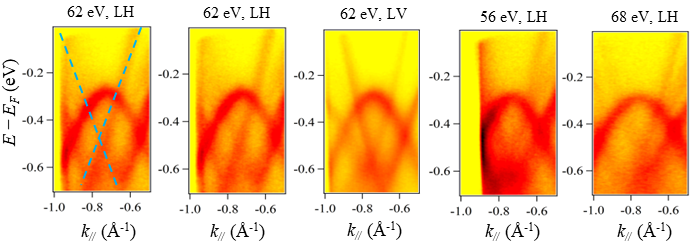} 
\caption{{Dirac-like dipsersion at the $\overline{\text{X}}$ high-symmetry point.} Zoomed view of the band dispersion around the $\overline{\text{X}}$ high-symmetry point. The blue (dashed line) serves as guides to the eyes for better visualization of the Dirac-like states at the $\overline{\text{X}}$ point. The experimental data were taken at the SSRL beamline 5-2 at a temperature of 15 K.}
\label{figS5}
\end{figure*}

\begin{figure*}
\centering
\includegraphics[width=10cm]{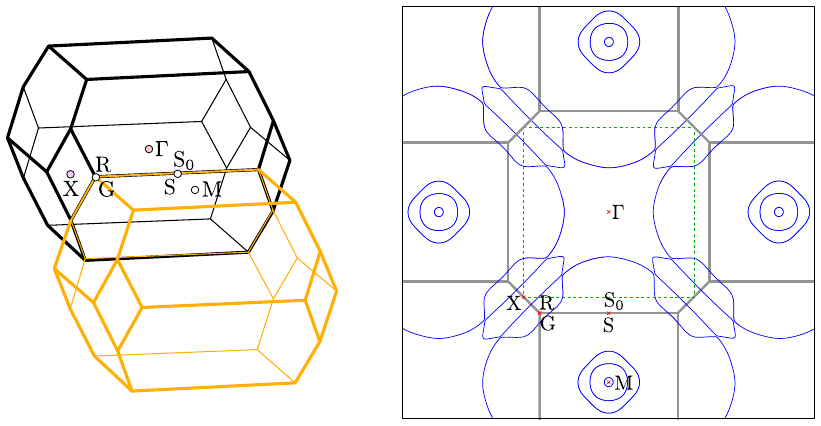}
\caption{{The two-dimensional bulk Fermi surface contour.}}
\label{figS11}
\end{figure*}
\end{center}
\end{document}